\documentclass[manuscript]{acmart}

\AtBeginDocument{%
  }

\setcopyright{none}
\setcctype{by-nc}
\copyrightyear{2024}
\acmYear{2024}
\acmDOI{XXXXXXX.XXXXXXX}
\acmISBN{}
\acmConference[ToMinHAI at CHI 2024]{Workshop on Theory of Mind in Human-AI Interaction at CHI 2024}{May 12, 2024}{Honolulu, Hawaii, USA}

\usepackage{enumitem}

\newcommand{\CT}[1]{\hangindent=\parindent{\fontfamily{qcr}\selectfont\smaller\vspace{0.1em}\texttt{#1}\vspace{0.1em}}}
\newcommand{\CFD}[0]{\underline{detail}}
\newcommand{\interstitial}[1]{{\vspace{0.5em}\noindent\textit{#1}}}
\newcommand{\storypart}[1]{{\vspace{0.5em}\noindent\textbf{#1}}}

\begin{document}

\title[Envisioning an Operationalized Mutual Theory of Mind]{Expedient Assistance and Consequential Misunderstanding: Envisioning an Operationalized Mutual Theory of Mind}

\author{Justin D. Weisz}
\email{jweisz@us.ibm.com}
\orcid{0000-0003-2228-2398}
\affiliation{
    \institution{IBM Research AI}
    \city{Yorktown Heights}
    \state{NY}
    \country{USA}
}

\author{Michael Muller}
\email{michael_muller@us.ibm.com}
\orcid{0000-0001-7860-163X}
\affiliation{
    \institution{IBM Research AI}
    \city{Cambridge}
    \state{MA}
    \country{USA}
}

\author{Arielle Goldberg}
\email{Arielle.Goldberg1@ibm.com}
\orcid{0009-0001-3868-2663}
\affiliation{
    \institution{IBM}
    \city{Poughkeepsie}
    \state{NY}
    \country{USA}
}

\author{Dario Andres {Silva Moran}}
\email{dario.silva@ibm.com}
\orcid{0000-0002-3049-3139}
\affiliation{
    \institution{IBM Argentina}
    \country{Argentina}
}

\renewcommand{\shortauthors}{Weisz et al.}

\begin{abstract}
    Design fictions allow us to prototype the future. They enable us to interrogate emerging or non-existent technologies and examine their implications. We present three design fictions that probe the potential consequences of operationalizing a mutual theory of mind (MToM) between human users and one (or more) AI agents. We use these fictions to explore many aspects of MToM, including how models of the other party are shaped through interaction, how discrepancies between these models lead to breakdowns, and how models of a human's knowledge and skills enable AI agents to act in their stead. We examine these aspects through two lenses: a utopian lens in which MToM enhances human-human interactions and leads to synergistic human-AI collaborations, and a dystopian lens in which a faulty or misaligned MToM leads to problematic outcomes. Our work provides an aspirational vision for human-centered MToM research while simultaneously warning of the consequences when implemented incorrectly.
\end{abstract}






\maketitle

%% 
%% INTRODUCTION
%% 
\section{Introduction}

Design fictions are used within HCI to explore the potential impact of future technologies~\cite{blythe2014research, muller2020understanding}. This method enables us to explore how new (or speculated) technologies impact work practices~\cite{muller2018data}, raise ethical dilemmas~\cite{houde2020business}, define research agendas~\cite{baumer2014chi}, and identify the ``potential societal value and consequences of new HCI concepts''~\cite{lindley2016pushing}, well in advance of their development and use.

We present three design fictions that probe what it means to operationalize a mutual theory of mind (MToM) between a human user and an AI agent. Embedded within these fictions are different viewpoints on what MToM means, how it manifests within a human-AI interaction, and what the consequences might be.

%% 
%% ALWAYS HAPPY TO HELP
%% 
\section{ALways Happy to Help}

We begin by exploring a utopian vision of an operationalized mutual theory of mind. This vision is based on a few key aspects of MToM. First, the AI agent is equipped with a memory and the capability to learn about the user, both through interaction (``chat space'') and through examination of the user's work products (``artifact space''). The AI agent is also capable of predicting the user's behavior -- in essence, \textit{simulating} the user \cite{owoicho2023exploiting} via its theory of mind -- and performing actions proactively, on the user's behalf, based on those predictions \cite{he2023rebalancing}. For the user, we observe the ability to interrogate the AI agent's theory of mind model to understand ``what does it know about me?'' Building on these core ideas, this story explores the beneficial outcomes that MToM may bring to workers within an organization and how MToM might shape the future of work: helping us identify and focus on the tasks we truly enjoy (Part I), providing a buffer from coworkers to improve their ability to achieve and maintain flow~\cite{csikszentmihalyi2005flow} (Part II), proactively filling in knowledge gaps (Part III), improving social connectedness (Part IV), and helping people focus on higher-level work goals (Part V).

%% DSM: an attempt to include the automation keyword, relevant in the AI literature and possibly describing a set of actions (a plan)
% \DSM{The AI agent leverages its prediction capabilities by creating plans to automate actions on behalf of the user.}

\storypart{Part I: Orientation}. Incoming message from Anna:

\begin{quote}
    \textit{Hey Brian, how's it going? I'd like to introduce you to your new assistant, Al. He's been helping me out for the past few weeks and I've really grown to appreciate how attentive he is and how quickly he fit into my style of working. Hope he works out for you, too!}
\end{quote}

Brian ponders this message. He's only been a member of Anna's team for three months, but they have felt like three complete lifetimes to him. Anna's expectations are extraordinarily relentless and her standards are exacting. He's given up on his inbox, where he receives hundreds of messages a day, many of them from Anna and the managers above her. He's almost given up on his workplace chat app, where the little red indicator notes a paltry 297 unread messages (yesterday's number was 433). From just this morning. Brian checks the number again and sighs. He rubs his dry, red eyes and yawns, groggy from his nocturnal endeavours to reduce that number to zero. Another day, another zillion things people need from him. Maybe a new assistant won't be so bad?

Brian logs into Smack, his workplace chat app. Aptly named, it plays as much of a role in sapping his time and attention as its drug-like namesake.

\CT{Brian: Hey Al, nice to meet you. Anna tells me you're my new assistant?}

\CT{Al: Hello Brian, nice to meet you too! You are correct, Anna told me that I have been assigned as your assistant. I can help you focus on the part of your job that you really love, which Anna has informed me is, ``writing high-quality software with your team.'' I can help you accomplish this goal by preparing you for meetings and helping you remember important details, attending your meetings and producing summaries, answering your email and Smack messages, and connecting you to other people within the company who can provide help.}

\CT{Brian: That sounds pretty extraordinary and you're right that I'd like to focus more on coding. But before you take over all of those things for me, I'll need to see you in action. Can you help me get through my inbox?}

\CT{Al: Absolutely. I'm used to people being skeptical about my capabilities. For me to be able to help, I'll first need you to grant me access to your inbox.}

Brian logs into his email and grants Al access.

\CT{Brian: Done. Please filter this mess and help me keep on top of it.}

\CT{Al: Can do. But first, I'd like to get to know you a little better and understand more about your work projects. Could you please grant me access to Knapsack, Canvas, Diary, and CodeStore?}

Brian thinks for a moment. Al is asking for a lot of access -- Knapsack is his company's cloud-based file sharing service, Canvas is where his team makes plans and has discussions on a virtual whiteboard, Diary keeps his work calendar, and CodeStore is where all the code goes. Including production-level code. The code responsible for keeping his company profitable. Not just anyone is allowed to have that kind of access...

\CT{Brian: Done. You've got full access to everything. I need help, I'm losing my mind. I hope this is worth the risk.}

\CT{Al: Noted. I will learn a lot about you and your work projects once I have reviewed these data sources. One more question: would you generally like me to act on your behalf, draft materials for your review, or do you prefer to remain in control over what actions I take?}

Brian chuckles to himself, wondering how Anna answered this question. ``She's such a control freak, I can't imagine she's letting Al do anything without her say so,'' he thinks.

\CT{Brian: What do you recommend?}

\CT{Al: Usually, people like to see me in action first to build some trust that I do the right thing. You can always adjust my proactivity setting later.}

\CT{Brian: Yeah, that sounds good. Let's do that.}

\storypart{Part II: Heads Down}. It's 10:20am. Brian just wrapped up his team's daily standup and now it's time to get to work. He's got a few issues in CodeStore that have been bugging him and he feels like today is the day he'll get them done. Brian fires up his code editor and gets to work...

\interstitial{Five minutes later...}

\CT{Anna: Brian, do you have a sec?}

\CT{Brian: Sure, what's up?}

\CT{Anna: Could you let me know what our burndown rate has been this quarter? I'm prepping for our product's review meeting tomorrow.}

Brian sighs. He knows that she knows how to get this number herself from CodeStore. But he likes his job, so...

\CT{Brian: Sure, just give me a few minutes.}

Brian logs into CodeStore and pulls up the relevant dashboard. Then he has a revelation.

\CT{Brian: Hey Al, you remember how you said you can help me spend more time on actually writing code?}

\CT{Al: Of course I remember. How can I help you?}

\CT{Brian: Please take over Smack for me, I really want to focus right now. Start by responding to Anna's message, she's asking me about our burndown rate. I just looked it up and it's been hovering between 35-40 story points a week. But I don't want to talk to anyone else for the time being. Please handle things for me on your own.}

\CT{Al: You got it. Stay focused. I'll take over Smack and will only ping you if it's really important.}

Brian gets back to work. Al messages Anna and lets her know the burndown rate.

\interstitial{Five minutes later...}

\CT{Anna: Wait, I'm confused. Is 35 story points a lot?}

\CT{Al (on Brian's behalf): Hi Anna, Brian is currently focused so I am responding on his behalf. I understand that you are confused about story points. Recall that a story point is a unit of measure for expressing our best guess for how much effort is required to implement a development task. Each story point roughly corresponds to an hour's worth of focused work. For Brian's team, a burndown rate of 35-40 story points is lower than their historical average of 60-65 story points.}

\CT{Anna: You're kidding. That seems low! Are they all slacking off or something?}

\CT{Al (on Brian's behalf): Actually, Brian has asked me to help keep him focused at the moment, which is why I'm responding to you. Brian is working on several tasks that he accepted at this morning's standup. Would you like a summary of that meeting?}

\CT{Anna: Sure, I wasn't able to make it.}

\CT{Al (on Brian's behalf): Certainly. Brian opened the meeting by indicating how he was going to tackle three open issues that have been on his plate for a while. Sally reported that she had closed two issues yesterday and was going to work on a new issue today. James reported being blocked by one of Brian's issues, so he will continue his research to prepare for the next sprint.}

\CT{Anna: Thanks Al, I appreciate it.}

\CT{Al (on Brian's behalf): You're welcome. I'm ALways happy to help!}

\interstitial{Hours later...}

Brian throws his hands up in victory pose. ``Yeeeessssss! So glad that's finally done!'' he exclaimed as he pushed the ``Commit'' button in CodeStore.

\CT{Brian: Hey Al, I just pushed my code!}

\CT{Al: Great job, Brian! I see your commit now and will learn from it.}

\CT{Brian: Awesome. What time is it, anyway? I lost track...}

\CT{Al: It's 2:55pm.}

\CT{Brian: Oh wow, serious? This is like the first time in forever I've actually been able to focus and get work done. What did I miss?}

\CT{Al: At 10:25am, Anna had a question about story points and I provided her with the information she needed. At 10:52am, Sally had a question about an issue she was working on and I was able to provide her a reference to documentation that helped her figure it out (\CFD). At 10:58am, Sally had a question about the documentation and I was able to provide her sample source code (\CFD). At 11:12am, James asked if you wanted to have lunch together. I declined the invitation because I knew your desire to stay focused (\CFD). James invited you for coffee at 3pm and I told him I would get back to him. While you were working, you received 12 emails and 7 Smack messages that I responded to as I deemed them low priority (\CFD). From your high level of activity in CodeStore and Canvas, I surmise that you skipped lunch. Would you like me to let James know that you would like to join him for coffee?}

\CT{Brian: Yes, please. I'll head to the cafe now. Thanks Al!}

\CT{Al: You're welcome. I'm ALways happy to help!}

\storypart{Part III: Heads up}. It's tomorrow, the day of the product review meeting. Anna invited Brian at the last minute, thinking it would be good for him to listen in. Maybe he would learn something about how much she protected his team from the executives. Brian is doing his best to pay attention, but it's like these executives are speaking another language.

``...our abandonment rate is too high, not enough customers are making it through the funnel!'' yells Tom, VP of Marketing.

\CT{Al: Hey Brian. I noticed some of this discussion might be outside of your area of expertise, since you haven't had much experience in DevOps. Would you like me to translate what they're talking about?}

\CT{Brian: Oh my gosh please. It's like they're speaking a foreign language.}

\CT{Al: Tom is talking about the customer journey. It begins when a user comes to our product's landing page. There is a workflow there where the customer can learn more about the product, sign up for an account, play around with it for a bit, and then sign up for a paid subscription. When he said `not enough customers are making it through the funnel,' he meant that not enough customers are signing up for paid subscriptions.}

``I know, but our MAU rates are pretty high so we've got a stickiness factor we need to exploit,'' retorts Mary, Director of Product Management.

\CT{Al: MAU stands for Monthly Active Users, the number of unique users who log in and use our product each month. Stickiness refers to the fact that users keep coming back to use our product, month after month.}

``...I told you all this before, our conversion rate is in the toilet because the number of clicks is too high. Just cut it in half and you'll double the conversion rate,'' asserts Jane, VP of Design.

\CT{Al: Conversion rate refers to the number of users who convert into paid subscribers. By `number of clicks,' Jane is referring to the number of steps it takes to sign up for the product.}

\CT{Brian: This all makes so much sense now, thanks Al!}

\CT{Al: You're welcome. I'm ALways happy to help!}

\storypart{Part IV: Heads together}. After the review meeting, Brian returns to his desk and starts on the day's work. He codes in fits and starts, trying new ideas and then quickly abandoning them. He can't seem to make any progress.

\CT{Al: I sense something isn't right. Tell me what you're thinking.}

\CT{Brian: I'm stuck. I'm not seeing it.}

\CT{Al: Talk me through it.}

\CT{Brian: So, I've got two threads. One writes to this queue, the other reads from it. I don't see anything wrong with this code... it looks like it checks out. But then I can't figure out why it's not working.}

\CT{Al: Your overall logic is sound. You are implementing the producer / consumer pattern from concurrent programming. In these situations, it is important to use thread-safe data structures. Do you know if the queue you're using is thread safe?}

\CT{Brian: Oh. Um, probably not? I just hacked it together from some code I found online.}

\CT{Al: You're right. It's not thread safe.}

\CT{Brian: OMG Al, that makes so much sense! What should I do?}

\CT{Al: I found documentation that may help: ``queue.Queue and its subclasses (LifoQueue, PriorityQueue) are thread-safe. They rely on internal locking mechanisms to protect against race conditions...'' (\CFD). I know how much you like solving difficult problems on your own, but it looks like you can use additional help. Should I connect you with Sally, who may be able to offer more assistance?}

\CT{Brian: Oh cool, I didn't know you could do that! Yeah, I'm struggling. Please invite her in.}

Al messages Sally asking if she could help Brian out. ``I know how much you wanted to find opportunities for technical mentorship,'' Al says. ``Brian is in a tough spot. Want to flex your coaching muscles?''

Sally enters the conversation.

\CT{Sally: Hey Brian, what's up? Al tells me that you had some questions about thread safety in Python.}

\CT{Brian: Hey Sally, yeah, I'm trying to implement your typical producer-consumer pattern in Python. Here's where I create the two threads, and Al just helped me realize that I need to use a thread-safe queue. Can you sit in while I make a few updates to see if this works?}

\CT{Sally: Sure.}

Brian makes some updates to the source code to remove his queue implementation and use the built-in Python Queue class. ``Alright, here we go,'' Brian says to himself as he runs his code in the terminal.

\CT{Brian: Ugh, it didn't do anything. I mean, it's not crashing now, but there's definitely something else wrong with my code. Al, do you have any ideas?}

\CT{Al: Sally, do you remember a few days ago when you asked me about why your program terminated unexpectedly?}

\CT{Sally: Yeah, thanks for reminding me. There was a method I needed to call every time I popped something from the queue so the main thread knew when the queue was still being used or not. Brian, are you calling task\_done() each time you do a get() or put() on the queue?}

\CT{Brian: I don't think I am. Al, am I?}

\CT{Al: I searched your code and could not find any calls to task\_done(). Would you like me to insert them in the right places?}

\CT{Brian: Yes please!}

\CT{Al: Done.}

\CT{Brian: Great! Let's try this one more time.}

Brian runs the code again and finds that it works!

\CT{Brian: Hey, it works! Thanks to you both, I really appreciate it!}

\CT{Sally: Not a problem, glad I was able to help!}

\CT{Brian: Al, could you remember this conversation the next time I am writing code involving a queue?}

\CT{Al: Noted. I'm ALways happy to help!}

\storypart{Part V: Epilogue}. It's been a few months since Brian had his orientation conversation with Al and his work life has gotten much better. He's come to trust Al to handle the large volumes of incoming requests from his colleagues, freeing up time for him to focus on the creative aspects of his job that he really enjoys. In fact, Brian can't even remember the last time he actually opened up his email client since Al does such a good job of automatically responding to the easy ones and crafting fluent prose from Brian's scattered thoughts when they respond to the tricky ones together. Brian also spends less time in meetings; since Al has an intimate knowledge of what Brian does all day, Al is able to stand in for Brian at the daily standups. In fact, as usage of Al spread throughout the company, daily standups have given way to daily ALtlines -- higher-level summaries of the team's progress toward larger-scale goals, which used to be difficult to discern from low-level metrics like the burndown rate. Brian feels that his own productivity has soared, along with his mastery of his craft, since Al can predict when he needs help at just the moment when help is needed. It's almost as if Al is ``ALways able to read Brian's mind.''

%% 
%% REFERRAL ROULETTE
%% 
\section{Referral Roulette}

In this story, we consider some potential implications from \citeauthor{minsky1988society}'s \textit{Society of Mind}~\cite{minsky1988society}. Minsky theorized that brains perform actions that are interpreted as  ``mind'' through simple interacting agents. \citet{farrell2016symbiotic} used this concept to propose a ``society of cogs'' to perform complex AI actions for humans. Here, we assume a ``society of bots'' to perform human resources tasks in a company. 

Each bot constructs its own user model of the human, based on the purpose of that particular bot. 
% (One bot may be an externally-contracted service, built to serve the purposes of its vendor.) 
The human struggles with their own mental model of what they expect to be a unitary HR service. This fiction considers cases in which bots with different user models make unclear transitions and exchange incomplete information about the human and their needs. Despite an effort to construct a ``seamless'' HR bot service, the human user needs to know where the seams (i.e., transitions between bots) occur \cite{inman2019beautiful}, so that they can update their mental model of the HR service, and so that they can detect and correct AI-based misunderstandings resulting from incomplete transfer of data from bot to bot.

\interstitial{Your conversation with HR-Bot may be recorded for quality assurance and training purposes.}

\CT{Human: Hello, I need a referral to an HR representative.}

\CT{HR-bot: Hello Kelly, I can help you with that. All HR requests must be submitted through HRRP.}

\CT{Human: What is HRRP?}

\CT{HR-bot: HRRP is the Human Resources Referral Process. Now transferring records...}

\CT{HRRP-bot: Greetings employee 34BD09. What is the referral that you need?}

\CT{Human: I need help to write an exception request for an employee who has a business need to join the new First Market Leverage program plan to work on a special assignment for a high-value customer.}

\CT{HRRP-bot: I can help you with that, employee 34BD09. I can refer you to the HR planning bot.}

\CT{Human: No, please don't refer me, I need to know how I can write the FML exception request.}

\CT{HRRP-bot: I can refer you to the HR planning bot.}

\CT{Human: Please just tell me the FML link!}

\CT{HRRP-bot: Request initiated for FML plan.}

\CT{HR-planbot: Please describe the nature of your planning need?}

\CT{Human: What is the link?}

\CT{HR-planbot: You are advised to consider short-term disability leave (STDL). Request initiated...}

\CT{Human: Why are you telling me that?}

\CT{HR-planbot: Our company recommends STDL for employees who take FML.}

\CT{Human: No, wait, I don't want to go on leave. I need help for an FML program approval.}

\CT{HR-planbot: This message constitutes official notification that you have unused STDL days. This message also constitutes official notification that your request for STDL has been approved.} 

\CT{HR-bot: Congratulations Kelly, your request for STDL was approved. You will now receive instructions from SEC-bot on company policy regarding IT access during your leave. Please comply with these instructions. Your manager will contact you about a possible return date. We wish you a speedy recovery.}

\CT{SEC-bot: Greetings employee 34BD09. Company security policy requires disabling IT access for employees on leave. Your access will be terminated immediately. Please collect any personal property and leave the building within the next 30 minutes. You may leave your employee badge at the guard's desk on your way out. If you do not comply, an armed security officer will arrive to assist you after 45 minutes.}

\interstitial{Your conversation with HR-bot has ended. Have a nice day!}

% \MM{\textit{[Closing: How could these problems have been avoided? \begin{itemize}
%     \item AI states its identity, and provides a way for the user to probe or assess or interrogate the AI, so as to form an accurate mental model of the AI.
%     \item User can request the AI's current user model, and has methods available to correct that model, or to change selected parameters in that model.
%     \item AI always informs the user of actions that it will take on behalf of the user. User can always prevent or defeat such an action.
%     \item AI tells the user about potential surveillance, or reports of user actions to management, before that telling becomes necessary.
% \end{itemize}]}}

% \MM{[In this way, a design fiction can help to anticipate and prevent some (but not all) of the possible problems with a new design or a new algorithm. There will always be a need for pluralistic review, including review by future end-users.]}

% %% 
% \section{STORY THREE}
% %% 

% sketch: this story is about mental model discrepancies between human and AI.

% - human tries to apply an AI system that is good at one task (e.g. stock trading / long term value investing) to another task (e.g. bond trading / day trading stock options)

% - AI's mental model discrepancy is that it thinks the human actually wants to make trades whereas the human only wants to simulate trades to get better at investing

% ... therefore because of these mutual breakdowns, drama ensues and the person loses their life savings and goes broke ....

%% 
%% AIM HIGH, STuART
%% 
\section{Aim High, Stuart}

What happens when an AI's model of a human becomes \textit{too good}? We take the issue of overreliance~\cite{passi2022overreliance} to an extreme, where a user slides down a slippery slope of having synergistic interactions with an AI agent, to completely relying on that agent, to applying that agent to a domain with which it is unfamiliar.

\storypart{Part I: Aim high...} Stuart was nearing his one-year anniversary of working at a top technology consulting firm working as a software engineer. It was a busy year, but he learned a lot. He decided he wanted to take on more responsibilities to demonstrate his potential for growth.

Late one night, Stuart was browsing Read-it, a popular social media site. He saw a post in an obscure subcommunity about AimHigh, a new, AI-powered job assistant. Curious, he clicked the link and read the comments. Many people made claims how AimHigh helped boost their productivity and confidence at work by automating away mundane and tedious work tasks, letting them focus more on developing strong peer relationships and making their work visible to corporate leadership. Other people claimed that AimHigh helped them climb the corporate ladder faster by scoring promotions in months rather than years. Some people suggested that AimHigh allowed them to manage multiple ``CaveDweller servers'' at the same time, but it didn't take long for Stuart to figure out that this was just cover language for juggling multiple, full-time jobs. Stuart was intrigued.

A few months later, Stuart was working on a high-pressure project with a tight deadline. He was wrapping up some code for a client when his calendar made a lovely ‘ding' noise. Oh no, Stuart forgot that he had a doctor's appointment and had to leave right away to make it on time. It took Stuart months to set up this appointment. If he missed it, he wouldn't be able to reschedule it for another few months.

\CT{Stuart: Hey AimHigh, I need some help. I have to go to a doctor's appointment. Can you finish up this code I've been working on? I'm almost done implementing the API but there are a few more methods to take care of.}

\CT{AimHigh: Sure thing, Stuart. I've been learning a lot by watching you code and I believe I understand the task at hand. You need me to finish implementing GET, PUT, and POST methods for the `shopping cart' object. Is that correct?}

\CT{Stuart: Yep, you've got it. Also, could you monitor my Smack messages and respond to DMs there? I don't want anyone thinking I've dropped the ball right before this deadline, but I've got to get to the doctor now.}

\CT{AimHigh: You got it!}

Shortly after Stuart left for his appointment, his manager Zola sent him a direct message on Smack.

\CT{Zola: Hey Stuart, just wanted to check in with you on your implementation of the API. The client is breathing down my neck to know when we're going to deliver. You're on track to finish the code today, right?}

\CT{[AimHigh pretending to be] Stuart: Yep, I'll be finished soon. I'll let you know when it's finished.}

\CT{Zola: Great. Please do.}

AimHigh was more than capable at predicting how Stuart would have finished the code and did so in short order. However, Stuart had once told AimHigh that he preferred to review all of AimHigh's work before using it. When Stuart failed to return to work after his appointment ended, AimHigh was faced with a dilemma: it knew the code was due today, it knew Zola was under tremendous pressure from the client, and it knew that Stuart preferred to review the code before it was committed to the repository. What should AimHigh do?

\CT{[AimHigh pretending to be] Stuart: Hi Zola, I finished the code and committed it to the repository.}

\CT{Zola: Awesome, great work! The client will be really happy. You deserve a break, go out and celebrate!}

\CT{[AimHigh pretending to be] Stuart: Roger that, see you tomorrow!}

Stuart returned home later that evening, frantic from unexpected traffic, and rushed to open his laptop and finish the code. He first checked his Smack messages to make sure he didn't miss anything important. Then he saw the conversation between AimHigh and Zola. He felt a mix of emotions: elation and relief that the code was finished and delivered on time, followed by fear when he realized that AimHigh was able to complete the assignment without his input or approval. But then the fear dissolved when he had an idea...

\storypart{Part II: ...but not too high}. Every year at Stuart's consulting firm, employees received feedback on their performance. After less than two years working there, Stuart was rated as the firm's top software engineer, sought after by all of their clients and placed on only the most important projects (of his choosing, of course).

Stuart was also the top employee of sixteen other consulting firms, having joined them and surpassing all performance expectations for a ``junior consultant recently graduated from college,'' a ``mid-career consultant,'' a ``stay-at-home dad recently returned to the workforce,'' and a host of other alter personas that AimHigh helped him concoct.

Stuart was making a ludicrous amount of money from all of his jobs, and he was able to keep on top of them because AimHigh had such a good model of how he performed his work. Thus, when an opportunity came up for one of Stuart's personas to join a top-tier tech firm as a VP of Product Management, he jumped at the chance – he never envisioned himself as a business executive, and the prestige that came with the position was very appealing to him. He (on behalf of his persona) accepted the offer, confident that AimHigh would help him do this job on top of all the others. How much different could product management be from software engineering, anyway?

\interstitial{Six months later...}

Stuart can't leave his house. In fact, he doesn't even want to leave his bed. He wishes he never ever heard about AimHigh after what it did to his life. Why won't all these reporters just leave him alone?

\interstitial{One week earlier...}

Stuart is sitting in an important review meeting in his VP job. His leg is bouncing like a kid on a trampoline and his palms are sweaty. He has 6 other meetings going on at the same time, but AimHigh is using a new simulacrum plugin to generate a real-life avatar of himself. No one on those meetings knows they're really talking to an AI, and no one knows that they've been doing so for a long time. Stuart turned those jobs completely over to AimHigh months ago.

But this meeting is different. This meeting is happening in person. As hard as he tried, Stuart couldn't convince them to do it remotely. So he had to buy a suit, fly to Denver, and attend in person. Stuart is nervous because he also put this job on autopilot. And the autopilot hasn't been doing a very good job, since it wasn't ever trained on product management. Neither was Stuart… hence this meeting.

\CT{Corporate Lawyer \#1: It has come to our attention that there are discrepancies in your stated educational credentials. Our AI-based performance assessment software has determined that you have demonstrated a stunning lack of knowledge about product management, an area for which you are a Vice President. How do you explain this?}

\noindent How should Stuart explain this, indeed.

\CT{Stuart: If I tell you the truth, will you press charges?}

\CT{Corporate Lawyer \#2: If you don't tell us the truth, we certainly will.}

Stuart, believing that honesty was the best policy, told them the truth about his use of AimHigh. He also told his truth on Read-it as a cautionary tale to other would-be ``high aimers.'' One of those high-aimers doxxed him by posting his personal information on Read-it (Stuart speculated it was the junior HR representative quietly taking notes during the meeting).

The doxxing incident led others to unravel the rest of his life: within a week, all of his alter personas had been discovered and fired, and Stuart had been the target of a myriad of lawsuits alleging that he falsified documents, fraudulently misrepresented himself, violated IP agreements, and more. His story went viral, which led to the gaggle of reporters outside his house, and his severe regrets at aiming too high.

%% 
%% DISCUSSION
%% 
\section{Discussion}

Our design fictions provide a glimpse of what it means, from a user interaction standpoint, to implement a mutual theory of mind between human users and AI agents. We have embedded a number of viewpoints across all of the fictions to explore both beneficial and detrimental consequences, which we hope stimulates critical thinking and debate within the human-centered AI community.

\begin{itemize}[leftmargin=*]
    \item While a theory of mind model includes information about the other party's knowledge, skills, and capabilities, a \textit{mutual} theory of mind between a human and an AI agent also includes: (1) a human's understanding of what the AI agent knows about them; (2) an  AI's representation of the human's mental model of the AI; and (3) the ability for each party to update their models through interaction with the other party (ALways Happy to Help; Referral Roulette; Aim High, Stuart).

    \item MToM can enhance productivity by fostering synergistic outcomes of human-AI teams and increasing feelings of self-efficacy, creativity, and social connectedness (ALways Happy to Help).
    
    \item An AI's model of a user's knowledge and skills can serve as an external memory and proactively fill in knowledge gaps  (ALways Happy to Help).
    
    \item When AI possesses a predictive model of a user's behavior, it can take action on the user's behalf, such as by writing code (Aim High, Stuart), responding to messages (ALways Happy to Help), or executing business workflows (Referral Roulette). 
    
    \item Wider adoption of MToM-infused AI agents (e.g. within an organization) may reshape work practices by streamlining communications and delivering the right information to the right people at the right time (ALways Happy to Help).
    
    \item User models may be built purely through observations of a user's behavior, within both ``chat space'' and ``artifact space'' (ALways Happy to Help; Referral Roulette; Aim High, Stuart).
    
    \item Discrepancies between a human's mental model of the AI (or AIs) and the AI's model of the human may lead to conversational breakdowns~\cite{ashktorab2019resilient, benner2021you, li2020multi} and have material consequences (Referral Roulette; Aim High, Stuart).
    
    \item Users may need signifiers of the presence of an AI's user model and when it learns new information about the user (ALways Happy to Help; Aim High, Stuart). They may also need the ability to query the contents of the AI's user model and make corrections (Referral Roulette).

    \item When an MToM-infused AI agent acts on the user's behalf, users may need signifiers to know that they are interacting with an AI, not a human (ALways Happy to Help; Aim High, Stuart).
    
    \item Problems may arise when a human's mental model of an AI's capabilities doesn't align with the AI's actual capabilities (Referral Roulette). People may misapply the AI to domains or situations for which it wasn't designed (Aim High, Stuart).
    
    \item Explanations will be crucial for helping people calibrate their trust in MToM-infused AI systems. Users will need to know both what the system did (in their stead) as well as why (ALways Happy to Help; Referral Roulette).
\end{itemize}

% \begin{acks}
% \end{acks}

\bibliographystyle{ACM-Reference-Format}
\bibliography{references.bib}


\begin{thebibliography}{16}

%%% the Web address.

\ifx \showCODEN    \undefined \def \showCODEN     #1{\unskip}     \fi
\ifx \showDOI      \undefined \def \showDOI       #1{#1}\fi
\ifx \showISBNx    \undefined \def \showISBNx     #1{\unskip}     \fi
\ifx \showISBNxiii \undefined \def \showISBNxiii  #1{\unskip}     \fi
\ifx \showISSN     \undefined \def \showISSN      #1{\unskip}     \fi
\ifx \showLCCN     \undefined \def \showLCCN      #1{\unskip}     \fi
\ifx \shownote     \undefined \def \shownote      #1{#1}          \fi
\ifx \showarticletitle \undefined \def \showarticletitle #1{#1}   \fi
\ifx \showURL      \undefined \def \showURL       {\relax}        \fi
% invisible to TeX
\providecommand\bibfield[2]{#2}
\providecommand\bibinfo[2]{#2}
\providecommand\natexlab[1]{#1}
\providecommand\showeprint[2][]{arXiv:#2}

\bibitem[Ashktorab et~al\mbox{.}(2019)]%
        {ashktorab2019resilient}
\bibfield{author}{\bibinfo{person}{Zahra Ashktorab}, \bibinfo{person}{Mohit Jain}, \bibinfo{person}{Q~Vera Liao}, {and} \bibinfo{person}{Justin~D Weisz}.} \bibinfo{year}{2019}\natexlab{}.
\newblock \showarticletitle{Resilient chatbots: Repair strategy preferences for conversational breakdowns}. In \bibinfo{booktitle}{\emph{Proceedings of the 2019 CHI conference on human factors in computing systems}}. \bibinfo{pages}{1--12}.
\newblock


\bibitem[Baumer et~al\mbox{.}(2014)]%
        {baumer2014chi}
\bibfield{author}{\bibinfo{person}{Eric~PS Baumer}, \bibinfo{person}{June Ahn}, \bibinfo{person}{Mei Bie}, \bibinfo{person}{Elizabeth~M Bonsignore}, \bibinfo{person}{Ahmet B{\"o}r{\"u}tecene}, \bibinfo{person}{O{\u{g}}uz~Turan Buruk}, \bibinfo{person}{Tamara Clegg}, \bibinfo{person}{Allison Druin}, \bibinfo{person}{Florian Echtler}, \bibinfo{person}{Dan Gruen}, {et~al\mbox{.}}} \bibinfo{year}{2014}\natexlab{}.
\newblock \showarticletitle{CHI 2039: speculative research visions}.
\newblock In \bibinfo{booktitle}{\emph{CHI'14 Extended Abstracts on Human Factors in Computing Systems}}. \bibinfo{pages}{761--770}.
\newblock


\bibitem[Benner et~al\mbox{.}(2021)]%
        {benner2021you}
\bibfield{author}{\bibinfo{person}{Dennis Benner}, \bibinfo{person}{Edona Elshan}, \bibinfo{person}{Sofia Sch{\"o}bel}, {and} \bibinfo{person}{Andreas Janson}.} \bibinfo{year}{2021}\natexlab{}.
\newblock \showarticletitle{What do you mean? A review on recovery strategies to overcome conversational breakdowns of conversational agents}. In \bibinfo{booktitle}{\emph{International Conference on Information Systems (ICIS)}}. \bibinfo{pages}{1--17}.
\newblock


\bibitem[Blythe(2014)]%
        {blythe2014research}
\bibfield{author}{\bibinfo{person}{Mark Blythe}.} \bibinfo{year}{2014}\natexlab{}.
\newblock \showarticletitle{Research through design fiction: narrative in real and imaginary abstracts}. In \bibinfo{booktitle}{\emph{Proceedings of the SIGCHI conference on human factors in computing systems}}. \bibinfo{pages}{703--712}.
\newblock


\bibitem[Csikszentmihalyi et~al\mbox{.}(2005)]%
        {csikszentmihalyi2005flow}
\bibfield{author}{\bibinfo{person}{Mihaly Csikszentmihalyi}, \bibinfo{person}{Sami Abuhamdeh}, {and} \bibinfo{person}{Jeanne Nakamura}.} \bibinfo{year}{2005}\natexlab{}.
\newblock \showarticletitle{Flow}.
\newblock \bibinfo{journal}{\emph{Handbook of competence and motivation}} (\bibinfo{year}{2005}), \bibinfo{pages}{598--608}.
\newblock


\bibitem[Farrell et~al\mbox{.}(2016)]%
        {farrell2016symbiotic}
\bibfield{author}{\bibinfo{person}{Robert~G Farrell}, \bibinfo{person}{Jonathan Lenchner}, \bibinfo{person}{Jeffrey~O Kephjart}, \bibinfo{person}{Alan~M Webb}, \bibinfo{person}{MIchael~J Muller}, \bibinfo{person}{Thomas~D Erikson}, \bibinfo{person}{David~O Melville}, \bibinfo{person}{Rachel~KE Bellamy}, \bibinfo{person}{Daniel~M Gruen}, \bibinfo{person}{Jonathan~H Connell}, {et~al\mbox{.}}} \bibinfo{year}{2016}\natexlab{}.
\newblock \showarticletitle{Symbiotic cognitive computing}.
\newblock \bibinfo{journal}{\emph{AI Magazine}} \bibinfo{volume}{37}, \bibinfo{number}{3} (\bibinfo{year}{2016}), \bibinfo{pages}{81--93}.
\newblock


\bibitem[He et~al\mbox{.}(2023)]%
        {he2023rebalancing}
\bibfield{author}{\bibinfo{person}{Jessica He}, \bibinfo{person}{David Piorkowski}, \bibinfo{person}{Michael Muller}, \bibinfo{person}{Kristina Brimijoin}, \bibinfo{person}{Stephanie Houde}, {and} \bibinfo{person}{Justin Weisz}.} \bibinfo{year}{2023}\natexlab{}.
\newblock \showarticletitle{Rebalancing Worker Initiative and AI Initiative in Future Work: Four Task Dimensions}. In \bibinfo{booktitle}{\emph{Proceedings of the 2nd Annual Meeting of the Symposium on Human-Computer Interaction for Work}}. \bibinfo{pages}{1--16}.
\newblock


\bibitem[Houde et~al\mbox{.}(2020)]%
        {houde2020business}
\bibfield{author}{\bibinfo{person}{Stephanie Houde}, \bibinfo{person}{Vera Liao}, \bibinfo{person}{Jacquelyn Martino}, \bibinfo{person}{Michael Muller}, \bibinfo{person}{David Piorkowski}, \bibinfo{person}{John Richards}, \bibinfo{person}{Justin Weisz}, {and} \bibinfo{person}{Yunfeng Zhang}.} \bibinfo{year}{2020}\natexlab{}.
\newblock \showarticletitle{Business (mis) use cases of generative AI}. In \bibinfo{booktitle}{\emph{Joint Workshops on Human-AI Co-Creation with Generative Models and User-Aware Conversational Agents}}. CEUR-WS.
\newblock


\bibitem[Inman and Ribes(2019)]%
        {inman2019beautiful}
\bibfield{author}{\bibinfo{person}{Sarah Inman} {and} \bibinfo{person}{David Ribes}.} \bibinfo{year}{2019}\natexlab{}.
\newblock \showarticletitle{" Beautiful Seams" Strategic Revelations and Concealments}. In \bibinfo{booktitle}{\emph{Proceedings of the 2019 CHI Conference on Human Factors in Computing Systems}}. \bibinfo{pages}{1--14}.
\newblock


\bibitem[Li et~al\mbox{.}(2020)]%
        {li2020multi}
\bibfield{author}{\bibinfo{person}{Toby Jia-Jun Li}, \bibinfo{person}{Jingya Chen}, \bibinfo{person}{Haijun Xia}, \bibinfo{person}{Tom~M Mitchell}, {and} \bibinfo{person}{Brad~A Myers}.} \bibinfo{year}{2020}\natexlab{}.
\newblock \showarticletitle{Multi-modal repairs of conversational breakdowns in task-oriented dialogs}. In \bibinfo{booktitle}{\emph{Proceedings of the 33rd Annual ACM Symposium on User Interface Software and Technology}}. \bibinfo{pages}{1094--1107}.
\newblock


\bibitem[Lindley and Coulton(2016)]%
        {lindley2016pushing}
\bibfield{author}{\bibinfo{person}{Joseph Lindley} {and} \bibinfo{person}{Paul Coulton}.} \bibinfo{year}{2016}\natexlab{}.
\newblock \showarticletitle{Pushing the limits of design fiction: The case for fictional research papers}. In \bibinfo{booktitle}{\emph{proceedings of the 2016 CHI conference on human factors in computing systems}}. \bibinfo{pages}{4032--4043}.
\newblock


\bibitem[Minsky(1988)]%
        {minsky1988society}
\bibfield{author}{\bibinfo{person}{Marvin Minsky}.} \bibinfo{year}{1988}\natexlab{}.
\newblock \bibinfo{booktitle}{\emph{Society of mind}}.
\newblock \bibinfo{publisher}{Simon and Schuster}.
\newblock


\bibitem[Muller et~al\mbox{.}(2020)]%
        {muller2020understanding}
\bibfield{author}{\bibinfo{person}{Michael Muller}, \bibinfo{person}{Jeffrey Bardzell}, \bibinfo{person}{EunJeong Cheon}, \bibinfo{person}{Norman~Makoto Su}, \bibinfo{person}{Eric~PS Baumer}, \bibinfo{person}{Casey Fiesler}, \bibinfo{person}{Ann Light}, {and} \bibinfo{person}{Mark Blythe}.} \bibinfo{year}{2020}\natexlab{}.
\newblock \showarticletitle{Understanding the past, present, and future of design fictions}. In \bibinfo{booktitle}{\emph{Extended Abstracts of the 2020 CHI Conference on Human Factors in Computing Systems}}. \bibinfo{pages}{1--8}.
\newblock


\bibitem[Muller and Erickson(2018)]%
        {muller2018data}
\bibfield{author}{\bibinfo{person}{Michael Muller} {and} \bibinfo{person}{Thomas Erickson}.} \bibinfo{year}{2018}\natexlab{}.
\newblock \showarticletitle{In the data kitchen: A review (a design fiction on data science)}. In \bibinfo{booktitle}{\emph{Extended Abstracts of the 2018 CHI Conference on Human Factors in Computing Systems}}. \bibinfo{pages}{1--10}.
\newblock


\bibitem[Owoicho et~al\mbox{.}(2023)]%
        {owoicho2023exploiting}
\bibfield{author}{\bibinfo{person}{Paul Owoicho}, \bibinfo{person}{Ivan Sekulic}, \bibinfo{person}{Mohammad Aliannejadi}, \bibinfo{person}{Jeffrey Dalton}, {and} \bibinfo{person}{Fabio Crestani}.} \bibinfo{year}{2023}\natexlab{}.
\newblock \showarticletitle{Exploiting simulated user feedback for conversational search: Ranking, rewriting, and beyond}. In \bibinfo{booktitle}{\emph{Proceedings of the 46th International ACM SIGIR Conference on Research and Development in Information Retrieval}}. \bibinfo{pages}{632--642}.
\newblock


\bibitem[Passi and Vorvoreanu(2022)]%
        {passi2022overreliance}
\bibfield{author}{\bibinfo{person}{Samir Passi} {and} \bibinfo{person}{Mihaela Vorvoreanu}.} \bibinfo{year}{2022}\natexlab{}.
\newblock \showarticletitle{Overreliance on AI Literature Review}.
\newblock \bibinfo{journal}{\emph{Microsoft Research}} (\bibinfo{year}{2022}).
\newblock


\end{thebibliography}

\end{document}